\documentclass{article}

\usepackage{PRIMEarxiv}
\usepackage{url}

\usepackage[utf8]{inputenc} 
\usepackage[T1]{fontenc}   
\usepackage{hyperref}       
\usepackage{url}            % simple URL typesetting
\usepackage{booktabs}       % professional-quality tables
\usepackage{amsfonts}       % blackboard math symbols
\usepackage{nicefrac}       % compact symbols for 1/2, etc.
\usepackage{microtype}      % microtypography
\usepackage{lipsum}
\usepackage{fancyhdr}       % header
\usepackage{graphicx}       % graphics
\graphicspath{{media/}}     % organize your images and other figures under media/ folder
\usepackage{amsmath}
\usepackage{amssymb} 
\allowdisplaybreaks 
\title{Research on CNN-BiLSTM Network Traffic Anomaly Detection Model Based on MindSpore
}

\author{
  Qiuyan Xiang \\
  School of Information Technology \\
  Guangxi Police College   \\
  530028,China \\
  \texttt{3310263026@qq.com} \\
   \And
  Shuang Wu \\
  College of Computer Science and AI \\
  Southwest Minzu University   \\
  610213,China \\
  \texttt{ws18774555387@outlook.com} \\
   \And
  Dongze Wu \\
  Institute of Software \\
  Chinese Academy of Sciences   \\
  100190,China \\
  \texttt{dongze@isrc.iscas.ac.cn} \\
  \And
  Yuxin Liu \\
  School of Information Technology \\
  Guangxi Police College   \\
  530028,China \\
  \texttt{3105472417@qq.com} 
  \And
  Zhenkai Qin \\
  School of Information Technology \\
  Guangxi Police College   \\
  530028,China \\
  \texttt{qinzhenkai@gxjcxy.edu.cn} \\
}

\begin{document}
\maketitle

\begin{abstract}
% \lipsum[1]
With the widespread adoption of the Internet of Things (IoT) and Industrial IoT (IIoT) technologies, network architectures have become increasingly complex, and the volume of traffic has grown substantially. This evolution poses significant challenges to traditional security mechanisms, particularly in detecting high-frequency, diverse, and highly covert network attacks. To address these challenges, this study proposes a novel network traffic anomaly detection model that integrates a Convolutional Neural Network (CNN) with a Bidirectional Long Short-Term Memory (BiLSTM) network, implemented on the MindSpore framework. Comprehensive experiments were conducted using the NF-BoT-IoT dataset. The results demonstrate that the proposed model achieves 99\% across accuracy, precision, recall, and F1-score, indicating its strong performance and robustness in network intrusion detection tasks.
\end{abstract}

% keywords can be removed

\section{Introduction}
With the accelerated development of the Internet of Things (IoT) and Industrial Internet of Things (IIoT) technologies, the network structure is becoming increasingly complex. Survey data shows that the number of global Internet users has reached 5.5 billion in 2024 \cite{r1} , the number of cyber-attacks has increased by 28\% year-on-year \cite{r2} , and the scale of network traffic continues to climb, a change that poses a serious challenge to traditional security protection mechanisms. Timely detection of these anomalies is essential to ensure quality of service, avoid financial losses and maintain strong security standards \cite{r3} . Network traffic data usually consists of logs that summarise the communication between network-connected devices \cite{r4} , which contain a large amount of sensitive communication content and access patterns that, once maliciously accessed, can lead to information leakage or privilege abuse issues. In recent years, thanks to the continuous evolution of machine learning and deep learning technologies, data-driven network traffic anomaly detection methods have gradually become the focus of research. Among these approaches, deep learning models—such as Convolutional Neural Networks (CNN) \cite{r5}, Recurrent Neural Networks (RNNs) \cite{r6}, and their hybrid architectures—have markedly enhanced the accuracy and response efficiency of detection systems, owing to their superior capabilities in feature extraction and temporal sequence modeling.

In this experiment, we construct a network traffic anomaly detection model using a convolutional neural network (CNN) and a bidirectional long short-term memory network (BiLSTM) based on the MindSpore framework. First, the NF-BoT-IoT dataset is loaded, feature pre-processed, time series reconstructed and divided into training, validation and test sets. Then, the model structure is designed, the corresponding loss function and optimiser are configured, multiple rounds of training are performed, and the classification threshold is dynamically adjusted based on the performance of the validation set during the training process to improve the F1 score. After the type training is completed, Upon completing the type training, we load the parameters that exhibit optimal performance, evaluate them on the test set, and output the prediction results for further analysis. The experimental results demonstrate that the method achieves 99\% in key metrics such as accuracy, precision, recall, and F1 score, thereby confirming the effectiveness and robustness of the proposed model in the network intrusion detection task.

\section{Related Work}
With the development of the Internet, the massification of devices leads to the explosive growth of Internet traffic, which poses a major challenge to the management of network resources and the guarantee of network security \cite{r7}. In this context, network traffic anomaly detection, as an important means of identifying potential threats and abnormal behaviours in the network, has gradually become one of the core technologies for safeguarding the security and stable operation of computer networks. This technology is mainly used to discover abnormal communications or attacks in a timely manner by analysing the statistical characteristics and behavioural patterns in network traffic data. In recent years, thanks to the wide application of machine learning technology, network traffic anomaly detection gradually realises the transformation from rule-driven to data-driven, especially the intelligent methods represented by supervised learning and unsupervised learning, which greatly improve the accuracy and efficiency of detection \cite{r8}. In response to the complex and changing network security threats, researchers have continued to explore deeply in this field and proposed a variety of effective detection algorithms and model architectures, thus continuously promoting the network traffic anomaly detection technology in the direction of more efficient and smarter development.

During the early development of network traffic anomaly detection, research has mostly focused on statistical analysis and feature extraction methods, e.g., Lv et al \cite{r9} introduced the Wavelet Generalised Likelihood Ratio (WGLR) algorithm and Error Performance Detection (EPD) algorithm, which combine the wavelet transform and generalised likelihood ratio methods in order to capture fault points in real time. huang et al \cite{r4} proposed a Growth Hierarchy Based Self-Organising Mapping (GHSOM) anomaly detection method to understand anomalous network traffic behaviour and provide effective classification rules.Novakov et al \cite{r10} investigated the application of PCA and wavelet algorithms in network traffic anomaly detection and contributed significantly to the technological advancement in the field.Ding et al \cite{r11} demonstrated the use of Principal Component Analysis (PCA) in the detection of network traffic anomalies by analysing the Traffic Matrix (TM) features. Component Analysis (PCA) in network traffic anomaly detection.Bhuyan et al \cite{r12} proposed a multi-step outlier-based anomaly detection method for network-wide traffic.

In recent years, researchers have been exploring network anomaly detection strategies centred on machine learning techniques, especially the application of deep neural networks, such as CNNs and RNNs, on model architectures, which significantly improve the detection accuracy and operational efficiency of the system. For example, Wei et al \cite{r13} introduced a deep learning-based hierarchical spatio-temporal feature learning approach, specifically combining convolutional neural networks (CNN) and recurrent neural networks (RNN) for cyber anomaly detection.Hao et al \cite{singh2021ml} proposed a hybrid statistical-machine learning model for real-time anomaly detection of industrial cyber-physical systems, combining the dynamic thresholding model based on SARIMA with long and short-term memory (LSTM). model combined with a Long Short-Term Memory (LSTM) model.Singh et al. provided an overview of various machine learning techniques for network traffic anomaly detection, discussing the advantages and disadvantages of different models and their accuracy levels. In addition, Liu et al \cite{r15} developed a real-time anomaly detection system based on convolutional neural networks for online packet extraction and analysis.
% % \label{sec:headings}

% % \lipsum[4] See Section \ref{sec:headings}.

% % % \subsection{Headings: second level}
% % \lipsum[5]
% \begin{equation}
% \xi _{ij}(t)=P(x_{t}=i,x_{t+1}=j|y,v,w;\theta)= {\frac {\alpha _{i}(t)a^{w_t}_{ij}\beta _{j}(t+1)b^{v_{t+1}}_{j}(y_{t+1})}{\sum _{i=1}^{N} \sum _{j=1}^{N} \alpha _{i}(t)a^{w_t}_{ij}\beta _{j}(t+1)b^{v_{t+1}}_{j}(y_{t+1})}}
% \end{equation}

% \subsubsection{Headings: third level}
% \lipsum[6]

% \paragraph{Paragraph}
% \lipsum[7]

\section{Model Description}
% \label{sec:others}
% \lipsum[8] \cite{kour2014real,kour2014fast} and see \cite{hadash2018estimate}.

% The documentation for \verb+natbib+ may be found at
% \begin{center}
%   \url{http://mirrors.ctan.org/macros/latex/contrib/natbib/natnotes.pdf}
% \end{center}
% Of note is the command \verb+\citet+, which produces citations
% appropriate for use in inline text.  For example,
% \begin{verbatim}
%    \citet{hasselmo} investigated\dots
% \end{verbatim}
% produces
% \begin{quote}
%   Hasselmo, et al.\ (1995) investigated\dots
% \end{quote}

\begin{figure}[htbp]
  \centering
  \includegraphics[width=1.0\linewidth]{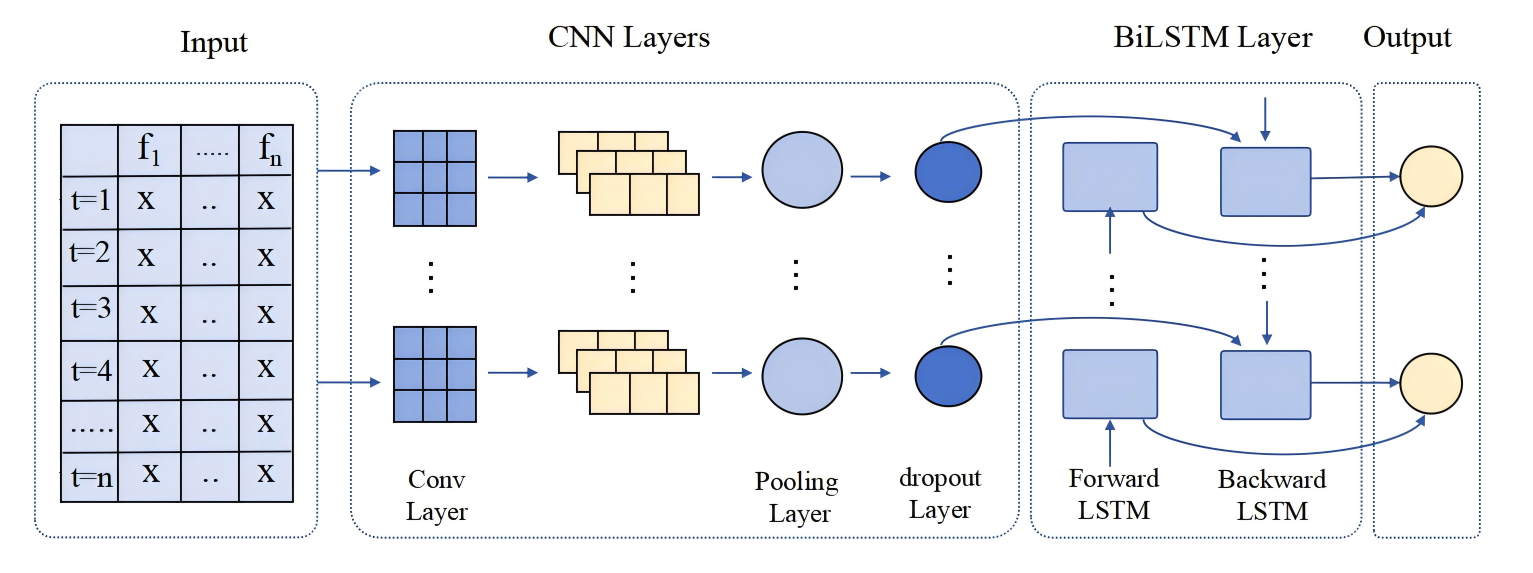}
  \caption{The overall architecture of the proposed CNN-BiLSTM network traffic anomaly detection model based on the MindSpore framework. The model integrates convolutional layers for local feature extraction and a bidirectional LSTM for temporal dependency modeling, enabling effective identification of complex spatiotemporal patterns in network traffic data.}
  \label{fig:model architecture}
\end{figure}

% \begin{center}
%   \url{https://www.ctan.org/pkg/booktabs}
% \end{center}
This model adopts the CNN-BiLSTM architecture, which combines the advantages of convolutional neural networks in local spatial feature extraction with the capability of bidirectional LSTM in temporal modelling. The specific model architecture is illustrated in Fig\ref{fig:model architecture}.
.In particular, the CNN layer mainly focuses on capturing spatial features in traffic data, while the BiLSTM layer further explores its temporal dependencies \cite{r16} to achieve collaborative modelling of spatial and temporal information. This structure is particularly suitable for analysing network traffic data under attack, and can effectively identify complex spatio-temporal feature patterns, thus improving the accuracy and robustness of abnormal behaviour detection. The model first takes the network traffic feature sequences in a fixed time window as input, extracts the local patterns through multi-layer 1D convolution and pooling, and prevents overfitting through a Dropout layer; then the convolution output sequences are fed into a BiLSTM layer, which captures the context-dependent information in the forward and backward directions, and then outputs the prediction results through the fully-connected layer and the activation function. The structure can take into account the spatial locality and temporal correlation of traffic data, and shows good performance and robustness in the anomaly detection task.
\subsection{CNN}
Convolutional neural networks (CNNs) were originally introduced by Yann LeCun et al. in 1989 \cite{r17}, aiming to process datasets with a grid structure, such as images and time-series data, through a weight sharing mechanism. In network traffic anomaly detection tasks, network flow data often contains rich local behavioural features, such as packet frequency and protocol fluctuations over a short period of time. By extracting traffic time slices frame by frame, CNN is able to identify potential anomaly patterns and improve the expressive power of pre-feature coding. In this experiment, a one-dimensional convolutional structure (Conv1D) is used to extract features from the reconstructed time series, and with the help of multi-layer convolution and pooling operations, the sensitivity of the model to local changes is enhanced to provide structured input for subsequent time-series modelling.

Given an input feature sequence matrix $\mathbf{X} \in \mathbb{R}^{T \times n}$, where $T$ denotes the number of time steps and $n$ represents the feature dimension at each time step, a one-dimensional convolution operation is applied by the CNN as follows:

\begin{equation}
y_i = \sum_{j=0}^{k-1} \mathbf{x}_{i+j} \cdot \mathbf{w}_j + \mathbf{b}
\end{equation}

where $\mathbf{x}_{i+j}$ denotes the feature at position $i+j$ in the input sequence; $\mathbf{w}_j$ is the convolution kernel parameter; $k$ is the kernel size;$\mathbf{b}$ is the bias term; and $y_i$ is the output feature after convolution.

\subsection{BiLSTM}
\begin{figure}[htbp]
  \centering
  \includegraphics[width=1.0\linewidth]{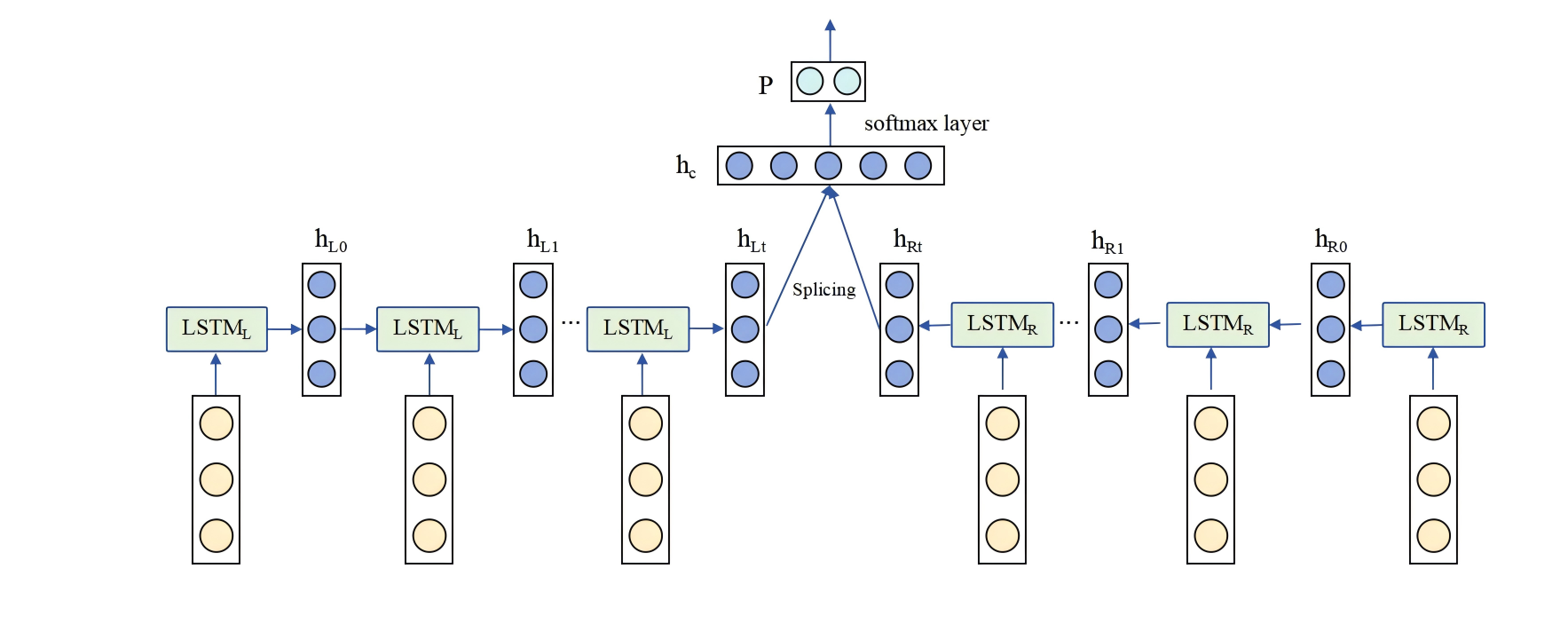} 
  \caption{Context-aware representation learning using a bidirectional LSTM. The left-to-right and right-to-left LSTM layers encode the input sequence into $\mathbf{h}{Lt}$ and $\mathbf{h}{Rt}$, which are concatenated to form the context vector $\mathbf{h}_c$. The final prediction $\mathbf{P}$ is obtained via a softmax layer.}
  \label{fig:BiLSTM model architecture}
\end{figure}

Bidirectional Long Short-Term Memory\cite{r18} was originally proposed by Alex Graves and Jürgen Schmidhuber in 2005, with the aim of considering both forward and backward contextual information in sequence modelling to improve the model's ability to model global sequence dependencies.LSTM has excellent long-term dependency modelling capabilities by introducing forgetting gates, LSTM has excellent long-term dependency modelling ability by introducing the mechanisms of forgetting gate, input gate and output gate, effectively retaining important historical information and suppressing irrelevant interference.The specific model architecture is illustrated in Fig\ref{fig:BiLSTM model architecture}. In this experiment, the BiLSTM network is used to jointly model the forward and backward time-series characteristics of network traffic, which improves the recognition ability and detection accuracy of abnormal behaviour.

First, the input sequence is processed in chronological order to compute the forward hidden state at each time step, which is used to extract forward temporal features:

\begin{equation}
\overrightarrow{h}_t = \text{LSTM}_L(x_t, \overrightarrow{h}_{t-1}), \quad t = 0, 1, \ldots, T
\end{equation}

Then, the input sequence is processed in reverse chronological order to compute the backward hidden state at each time step, which is used to extract backward temporal features:

\begin{equation}
\overleftarrow{h}_t = \text{LSTM}_R(x_t, \overleftarrow{h}_{t+1}), \quad t = T, T-1, \ldots, 0
\end{equation}

Finally, the forward and backward hidden states at each time step are concatenated to form a temporal representation that integrates past and future information:

\begin{equation}
h_c = \left[ \overrightarrow{h}_T ; \overleftarrow{h}_T \right]
\end{equation}

\noindent
where:
\begin{itemize}
    \item $t$ denotes the index of the current time step, where $0 \leq t \leq T$;
    \item $x_t$ represents the word embedding of the input sequence at time step $t$;
    \item $\overrightarrow{h_t}$ denotes the hidden state of the forward LSTM at time step $t$;
    \item $\overleftarrow{h_t}$ denotes the hidden state of the backward LSTM at time step $t$;
\end{itemize}

% \begin{itemize}
% \item Lorem ipsum dolor sit amet
% \item consectetur adipiscing elit. 
% \item Aliquam dignissim blandit est, in dictum tortor gravida eget. In ac rutrum magna.
% \end{itemize}
\section{Experimental Results and Analysis}
\subsection{Dataset}
The NF-BoT-IoT dataset\cite{r19} was meticulously developed by the University of Queensland based on the NetFlow format and represents a specialized dataset for Internet of Things (IoT) network security research. Its raw data originates from the BoT-IoT dataset, constructed by the Cyber Range Lab at the University of New South Wales Canberra (UNSW Canberra), Australia. The NF-BoT-IoT dataset comprises approximately 600,000 network flow records, among which only 2.31\% represent benign traffic, while a substantial 97.69\% correspond to malicious traffic, including various types of attacks such as reconnaissance, DDoS, DoS, and data theft. Owing to its comprehensive coverage of attack types and clearly labeled data, the dataset has been widely adopted as a standard benchmark in intrusion detection systems (IDS) and cybersecurity research. It provides a reliable and reproducible experimental platform, thereby significantly advancing the development of related studies.
\subsection{Experimental Environment}
The experiments were conducted on a local computing environment running Ubuntu 22.04. The hardware configuration includes an 8-core CPU and 32 GB of RAM. The software environment is based on Python 3.10, with core dependencies on the MindSpore deep learning framework, as well as common scientific computing libraries such as NumPy, pandas, and scikit-learn. The development platform is pre-installed with the ModelScope Library, supporting rapid development and seamless environment integration.
\subsection{Evaluation Metrics}
In this paper, Accuracy, Precision, Recall and F1 score are used as the main evaluation indexes of model performance. Among them, Accuracy is used to measure the proportion of samples on which the model's prediction results are consistent with the true labels, which is a basic indicator of the model's overall prediction ability. The higher the accuracy rate, the closer the model's classification results on all samples are to the real situation. Its specific calculation formula is as follows:

\begin{equation}
\text{Accuracy} = \frac{TP + TN}{TP + TN + FP + FN}
\end{equation}
\vspace{0.5em}
\begin{equation}
\text{Precision} = \frac{TP}{TP + FP}
\end{equation}
\vspace{0.5em}
\begin{equation}
\text{Recall} = \frac{TP}{TP + FN}
\end{equation}
\vspace{0.5em}
\begin{equation}
F1 = 2 \times \frac{\text{Precision} \times \text{Recall}}{\text{Precision} + \text{Recall}}
\end{equation}
\vspace{0.5em}

Among them, $TP$ denotes the number of positive samples correctly identified by the model, $TN$ represents the number of negative samples correctly identified, $FP$ refers to the number of negative samples incorrectly classified as positive, and $FN$ indicates the number of positive samples misclassified as negative. These metrics are used to evaluate the model's capability to distinguish between positive and negative classes.
\subsection{Experimental Results and Analysis}
\begin{figure}[htbp]
  \centering
  \includegraphics[width=0.8\linewidth]{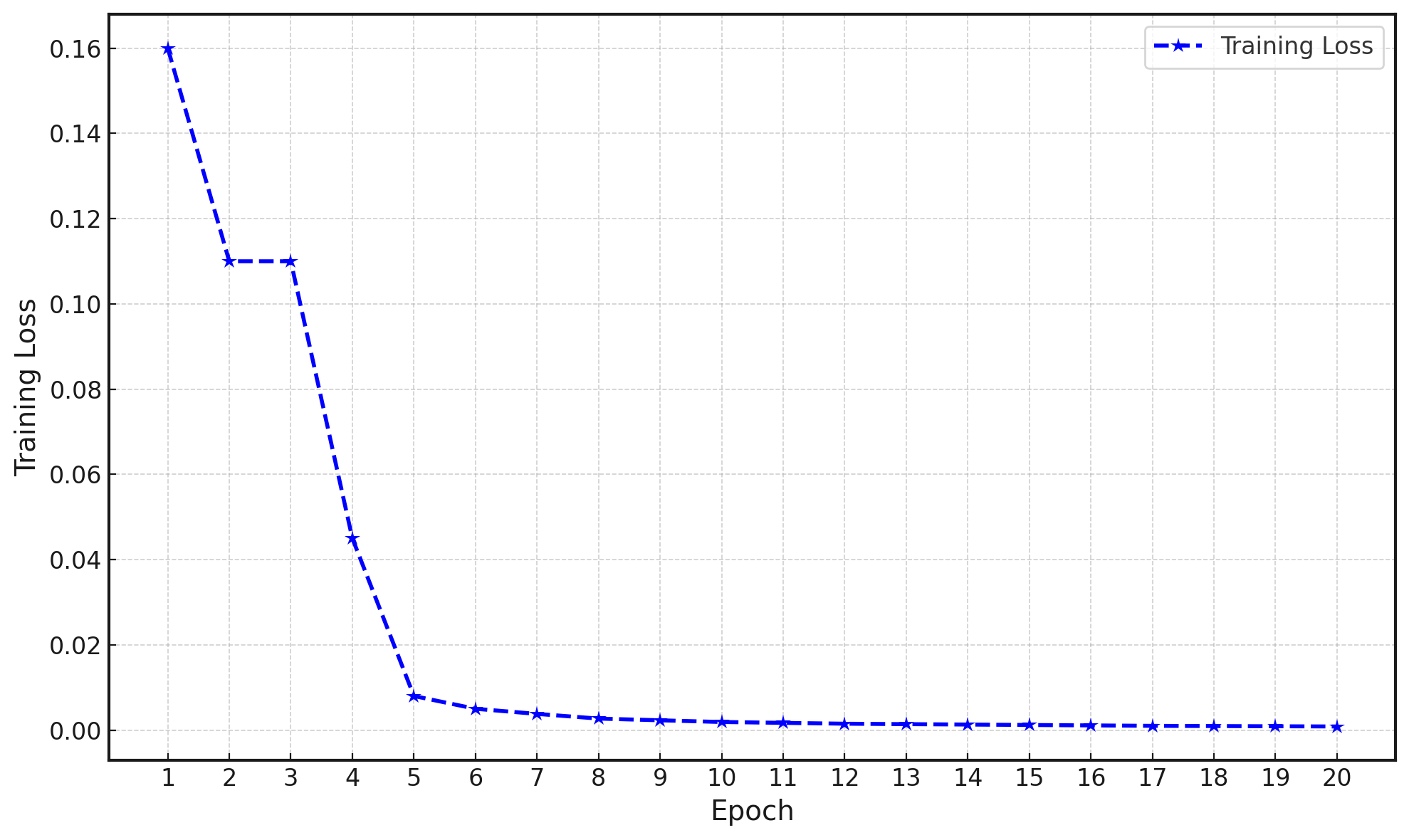} 
  \caption{Training loss curve of the CNN-BiLSTM model over 20 epochs. The model demonstrates rapid convergence and stable training dynamics, with the loss decreasing from 0.1650 to 0.0002.}
  \label{fig:training_loss}
\end{figure}

In this experiment, the model converges rapidly within 20 rounds of training, and the training loss is reduced from the initial 0.1650 to 0.0002, demonstrating good training stability and convergence, as shown in Fig\ref{fig:training_loss}. On the NF-BoT-IoT test set, the model exhibits extremely high classification performance with 99\% accuracy, precision, recall and F1 score, indicating that the model is able to efficiently identify anomalous behaviours in IoT traffic with a very low misclassification rate, which verifies the validity and robustness of the proposed method in real-world scenarios.

\section{Conclusion}
Aiming at the problem of insufficient detection accuracy of traditional security mechanisms in the complex network environment of IoT and industrial IoT, this paper constructs a network traffic anomaly detection model combining convolutional neural network (CNN) and bi-directional long and short-term memory network (BiLSTM) based on the MindSpore framework, and carries out systematic experiments on the NF-BoT-IoT dataset. The experimental results show that the accuracy, precision, recall and F1 score of the model on the NF-BoT-IoT dataset exceed 99\%, demonstrating excellent detection performance and strong generalisation ability. Although the model performs stably on a single dataset, it still faces challenges such as cross-domain adaptation and real-time optimisation in practical deployment, and future research can further combine techniques such as lightweight network structure and federated learning to improve the model's adaptability and deployment efficiency in multi-scenario environments.

\section*{Acknowledgments}
Thanks for the support provided by the MindSpore Community.
%Bibliography
\bibliographystyle{unsrt}  
\bibliography{references}

\end{document}